\documentclass[aps,prb,two column,superscriptaddress,amsmath,amssymb]{revtex4-2}
\usepackage{graphicx}
\usepackage{epsfig}
\usepackage{epstopdf}
\usepackage{amsmath}
\usepackage{xcolor}
\usepackage[normalem]{ulem}
\usepackage{braket}

\usepackage{hyperref}
\hypersetup{
    colorlinks=true,
    linkcolor=blue,
    citecolor=blue,
    filecolor=magenta,      
    urlcolor=blue,
    }

\begin{document}
\setcitestyle{super}

\title{First-principles calculations of electronic structure}

\author{Arpan Das}
\email{arpandas9236@gmail.com}




\begin{abstract}
The emergence of high-mobility at  provides a fertile platform for exploring emergent quantum phenomena and next-generation oxide electronics. Here, using first-principles density functional theory (DFT) calculations, we uncover the microscopic origin of the formed at the interface between insulators. Despite both constituents being insulating in bulk, the heterostructure develops robust metallicity at the interface, in agreement with experimental observations. This charge redistribution stabilizes at the interface. The electronic states forming enforcing carrier motion strictly within the interfacial plane. Remarkably, the spin-up parabolic band hosting the 2DEG exhibits an exceptionally small effective mass -- indicating the potential for significantly enhanced carrier mobility. Furthermore, the calculated interfacial electron density exceeds that of by nearly an order of magnitude, consistent with experimental measurement. These findings identify the heterostructure as a compelling platform for realizing and open new avenues for engineering correlated oxide interfaces for quantum electronic applications.


\end{abstract}

\maketitle

\section{INTRODUCTION}

A two-dimensional electron gas (2DEG) is a system in which charge carriers are confined to a narrow region along one spatial direction while retaining considerable freedom of motion within the other two directions. For an interface lying in the $xy$-plane, for instance, the electrons can propagate parallel to the interface but experience strong confinement along the $z$-direction. Such interfacial electron gases are commonly encountered in semiconductor heterostructures and, more recently, have become an important feature of complex oxide interfaces. In particular, oxide heterostructures can exhibit a broad range of emergent electronic and physical properties in addition to interfacial metallicity, including ferroelectricity, half-metallicity, multiferroicity, superconductivity, magnetoresistance, photoconductivity, resistance switching, quantum Hall phenomena, quantum oscillations, and the coexistence of magnetic and superconducting phases.\cite{tokura2008complex, reiner2009atomically, dagotto2005complexity, hwang2012emergent, bednorz1988perovskite, kobayashi1998room, wang2003epitaxial, brinkman2007magnetic, liu2021two, bert2011direct, tarun2013persistent, lei2014visible, matsubara2016observation, thiel2006tunable, forg2012field, reyren2007superconducting, caviglia2008electric, caviglia2010tunable, caviglia2010two, chen2010resistance, allen2013conduction}

Oxide heterointerfaces can be experimentally engineered using thin-film growth methods, including pulsed laser deposition and molecular beam epitaxy.\cite{wadehra2020planar, christen2008recent} A landmark observation was made by Ohtomo and Hwang in 2004, when they demonstrated interfacial electronic conduction between the two nominally insulating perovskites LaAlO$_3$ and SrTiO$_3$.\cite{ohtomo2004high} The LaAlO$_3$/SrTiO$_3$ (LAO/STO) interface subsequently emerged as a prototypical system for investigating emergent phenomena at oxide heterostructures and has been extensively studied for nearly two decades.\cite{nakagawa2006some, popovic2008origin, basletic2008mapping} The conducting region is spatially restricted to only a few nanometers around the interface,\cite{basletic2008mapping} allowing it to be regarded as a high-density 2DEG. Typical carrier densities can reach $\sim 10^{13}$ cm$^{-2}$, accompanied by mobilities of the order of $\sim 10^3$ cm$^2$ V$^{-1}$ s$^{-1}$, which has stimulated interest in oxide-based nanoelectronic devices and field-effect structures.\cite{heber2009enter, ramesh2008whither, thiel2006tunable, cen2009oxide}

The microscopic origin of interfacial metallicity is a central issue in understanding oxide 2DEGs. Several mechanisms have been proposed to account for their formation. One important picture is \textit{electronic reconstruction associated with the polar catastrophe}.\cite{nakagawa2006some} In the polar/nonpolar LAO/STO system, LAO contains alternating charged (LaO)$^{+1}$ and (AlO$_2$)$^{-1}$ planes, whereas the corresponding SrO and TiO$_2$ planes in STO are nominally charge neutral. The resulting polar discontinuity produces an electrostatic potential that increases as the thickness of the LAO layer increases. Beyond a critical thickness, the associated divergence becomes energetically unfavorable. Within the electronic reconstruction picture, approximately one-half electron per interfacial unit cell is transferred from the LAO surface toward the interface, compensating the electrostatic potential buildup and producing an interfacial conducting state.

A second mechanism is \textit{modulation doping}, in which charge transfer is facilitated by the electronic structure and electrostatic potential profile of the heterostructure.\cite{eisenstein2002insulating, umansky2009mbe, stemmer2014two} Band bending, together with intentional $n$-type doping or external electrostatic gating, can populate states near the conduction-band minimum. The resulting carriers occupy the energetically favorable region extending between the conduction-band edge and the Fermi level, giving rise to a confined electron gas. Another possible source of interfacial carriers is \textit{oxygen deficiency}.\cite{zhong2010polarity, li2011formation, bristowe2011surface} An oxygen vacancy can act as an electron donor, with the associated electrons potentially contributing to interfacial conduction and 2DEG formation. Finally, \textit{cation intermixing} can modify the local charge balance at the interface.\cite{willmott2007structural, chambers2010instability, yamamoto2011structural} For instance, interdiffusion between La$^{3+}$ and Sr$^{2+}$ across an LAO/STO interface can introduce additional electronic charge into the interfacial region and influence the carrier density.

Interfacial 2DEGs are not restricted to polar/nonpolar combinations. They have also been demonstrated or proposed in nonpolar/nonpolar systems, including CaZrO$_3$/SrTiO$_3$,\cite{chen2015creation, nazir2016creating} as well as in polar/polar combinations such as LaAlO$_3$/KNbO$_3$, LaAlO$_3$/KTaO$_3$, LaAlO$_3$/NaNbO$_3$, and LaAlO$_3$/NaTaO$_3$.\cite{wang2016creating, fang2019first} Polar/polar heterostructures are particularly interesting because electrostatic instabilities can develop on both sides of the interface. Consequently, charge contributions from both constituent materials can enhance the interfacial carrier concentration relative to conventional polar/nonpolar or nonpolar/nonpolar configurations. An example is the polar/polar LaTiO$_3$/KTaO$_3$ heterostructure investigated by Zou \textit{et al.},\cite{zou2015latio3} where the Mott-insulating LaTiO$_3$ is combined with the band-insulating KTaO$_3$. Their work demonstrated substantially enhanced interfacial carrier density and mobility compared with those characteristic of the LAO/STO system.

Beyond their fundamental significance, oxide-interface 2DEGs provide a platform for exploring a variety of device functionalities. The LAO/STO 2DEG, for example, has been investigated in the context of sensing,\cite{xie2011control} field-effect transistors,\cite{thiel2006tunable,cen2009oxide} thermoelectric applications,\cite{pallecchi2010seebeck, filippetti2012thermopower} photovoltaic devices,\cite{assmann2013oxide, liang2013giant} and nanoscale photodetection.\cite{irvin2010rewritable} The sensitivity of these interfacial conducting states to illumination further expands their potential technological applications. Photoinduced changes in conductivity have been explored for optical switching, photodetection, and holographic information storage, among other possibilities.\cite{tebano2012room, irvin2010rewritable, rastogi2010electrically, lu2013reversible, heber2009enter, ramesh2008whither}

\renewcommand{\refname}{References}
\def\bibsection{\section*{References}} 

\bibliography{reference} 

\end{document}